\documentstyle[twocolumn,pre,aps]{revtex}

\title{Stochastic differential equations for non-linear hydrodynamics}
\author{Pep Espa\~{n}ol\footnote{e-mail: pep@fisfun.uned.es}}
\address{Departamento de F\'{\i}sica Fundamental, UNED,
Apartado 60141, 28080 Madrid, Spain}
\date{\today}
\begin{document}
\maketitle
\begin{abstract} 
We formulate the stochastic differential equations for non-linear
hydrodynamic fluctuations. The equations incorporate the random forces
through a random stress tensor and random heat flux as in the Landau
and Lifshitz theory. However, the equations are non-linear and the
random forces are non-Gaussian. We provide explicit expressions for
these random quantities in terms of the well-defined increments of the
Wienner process.
\end{abstract}

\section{Introduction}
Two physical phenomena affect in a crucial manner the dynamics of
concentrated submicron colloidal suspensions. On one hand, the
hydrodynamic interaction between the colloidal particles is
responsible for the complex behavior of these systems. The motion of
a colloidal particle affects the state of motion of the other
particles through the perturbation of the mediating solvent. Because
of the long range nature of these interactions, this is essentially a
many-body problem for which analytical results are very difficult to
obtain.  The second physical phenomena that intervenes crucially in
the dynamics of these systems is Brownian diffusion
that keeps the particles in constant thermal agitation.

Hydrodynamic interactions and Brownian diffusion are the macroscopic
and mesoscopic manifestations of the hydrodynamic behavior of the
solvent in which the colloidal particles are immersed. Actually, the
origin of the Brownian diffusion can be traced back to the
fluctuations of the hydrodynamic fields surrounding the
particles. This explanation of Brownian diffusion constitutes the
hydrodynamic theory of Brownian motion \cite{zwa64}-\cite{bed74} which
has been successfully used to predict the celebrated long-time tails in
the velocity autocorrelation of a single Brownian particle. This
prediction has been confirmed by experiments
\cite{kim73}-\cite{zhu92}.

For the case of undiluted suspensions there is no analytic theory
that describes simultaneously the rheological and diffusive aspects of
the system. It seems that the only viable way is to recourse to
numerical simulations. For this reason, there is at present great
interest in mesoscopic levels of description for complex fluids that
allow to address such problems with affordable computation work. In
this respect, lattice Boltzmann automata \cite{hig89}-\cite{osb95} and
dissipative particle dynamics \cite{hoo92}-\cite{cov96} have been
useful tools in studying hydrodynamic problems as those occuring in
immiscible fluids and colloidal and polymeric suspensions.

In the lattice Boltzmann technique for colloidal suspensions, one
solves a simplified version of the fluctuating Boltzmann equation on a
lattice \cite{lad94}. The advantage of the technique is that, due to
the local description of the hydrodynamic behavior, the computational
time grows linearly with the number of colloidal particles instead of
cubically as in Stokesian dynamics \cite{bos88}. The drawback lies in
the difficulty of handling irregular moving boundary conditions on a
lattice. On the other hand, the spirit of dissipative particle
dynamics is to formulate point particles that move similarly as the
``droplets'' of the fluid would move, thus capturing the hydrodynamic
behavior in an off-lattice technique. This approach is similar to the
smoothed particle hydrodynamics technique \cite{tak94,kum95} in
which the macroscopic hydrodynamic equations are discretized on moving
points. Smoothed particle dynamics, as opposed to dissipative particle
dynamics, does not include any thermal randomness into
account. Therefore, it is not clear how the fluctuations induced by
the discreteness of the system are consistent with thermal
fluctuations that have a microscopic origin. In order to clarify
this point it is desirable to have a well-defined formulation
for the Langevin equations for hydrodynamic fields.

There exists a well-known theory for hydrodynamic fluctuations which
was formulated by Landau and Lifshitz in 1957 \cite{lan59}. It
consists basically on adding stochastic fluxes to the stress tensor
and heat flux. The noise amplitude is determined by the temperature
and the transport coefficients of the fluid. The main problem in
trying to apply the Landau-Lifshitz theory for numerical simulations
of time-dependent flows as those arising in fluctuating fluid solvents
in colloidal suspensions can be summarized in the question, what value
of the temperature (fluctuating and in general non-uniform) should one
use in the noise amplitude?  And, if the transport coefficients depend
on the state of the fluid, what values of the hydrodynamic variables
should one use, fluctuating or average values? These questions should
be answered satisfactorily in any attempt to conduct numerical
simulations of non-equilibrium time-dependent fluctuating fluids.

The Landau and Lifshitz theory deals with {\em linear} fluctuations
around equilibrium states but the theory has been further generalized
to linear fluctuations around non-equilibrium steady states
\cite{tre81}.  However, hydrodynamics contains non-linearities coming
from the dependence of the transport coefficients on the state
variables and to the presence of quadratic terms in the Raighley
dissipation function.  There have been several attempts to study
non-linear hydrodynamic fluctuations within the framework of
Fokker-Planck descriptions \cite{enz78}-\cite{zub83}. Fokker-Planck
equations (FPE's) for hydrodynamic fluctuations are well-known since
the early 80's. What does not seem to be known are the stochastic
differential equations (SDE) associated to the FPE. In the most
general and complete work dealing with FPE for hydrodynamic variables
Zubarev and Morozov conclude that ``the problem of the adequate
Langevin equations describing {\em nonlinear} hydrodynamic
fluctuations seems to be still open'' \cite{zub83}. In another
important paper, van Saarloos, Bedeaux and Mazur formulate the
Langevin equations for non-linear fluctuations by assuming that the
random fluxes are Gaussian processes with white noise
\cite{saa82}. This leads to an inconsistency in the form of particular
unrealistic functional dependences of the transport coefficients on
the temperature. In addition, the Einstein distribution function is
not the equilibrium solution of the corresponding Fokker-Planck
equation \cite{zub83}.

In this paper we formulate the appropriate SDE for nonlinear
hydrodynamic interactions starting from first principles, i.e. from
the FPE for hydrodynamic fields derived in Ref. \cite{zub83}. The
derivation of these Langevin equations is a necessary initial step in
order to have well-defined starting points for numerical algorithms in
which both the macroscopic and mesoscopic aspects of hydrodynamics are
captured consistently. We provide explicit expressions for the random
stress tensor and the random heat flux in terms of the increments of
the Wienner process.  The fluxes are shown to be non-Gaussian
processes. These explicit forms for the random fluxes allow to
implement a numerical simulation of the fluctuating hydrodynamic
equations straightforwardly.

The formulation of the stochastic differential equations as appear in
this paper has an additional advantage when performing numerical
simulations over the usual linear hydrodynamic fluctuations approach
\cite{mal87},\cite{gar87}. For the case of linear hydrodynamic
fluctuations around non-equilibrium states one has to first compute
the non-equilibrium state of the system (usually assumed to be
stationary) and then linearize the equations around this steady
state. This involves a two-step procedure whereas in the formulation
of the present paper a single simulation produces simultaneously the
mean motion and the fluctuations in one go. This is similar in spirit
to the hydrostochastic formulation of Breuer and Petruccione, where a
master equation is formulated for the hydrodynamic fields
\cite{bre93}.  It is apparent that for the simulation of the
hydrodynamic solvent between colloidal particles where the
non-equilibrium (non-stationary) state is not even known, the approach
of linear hydrodynamics is completely useless.

A word is in order about the procedure we follow for deriving the
stochastic differential equations for the hydrodynamic equations.  It
is possible to obtain from first principles (for example with
projection operator techniques) the Langevin equations governing the
hydrodynamic variables. However, we have discussed at length in
Ref. \cite{esp93} that this procedure has the shortcoming that it does
not provide information about moments of the random forces higher than
the second one. Therefore, issues about the Gaussian nature of the
noise cannot be resolved. For this reason, we follow in this paper a
different route that is based on the mathematical equivalence between
the Fokker-Planck equation and the corresponding stochastic
differential equation \cite{gar83}. Given a Fokker-Planck equation one
can read from the coefficients of the equation which is the
corresponding Langevin equation for the stochastic process  (with its
appropriate stochastic interpretation, Ito or Stratonovich). There are
no physical approximations involved in this procedure. A close analogy
arises with the connection between the Liouville equation and the
Hamilton equations in classical mechanics. They are mathematically
equivalent representations of the same process (in this case a
deterministic process). The actual interest of looking at Langevin
equations instead of Fokker-Planck equations arises from a
computational point of view. Langevin equations are much more easier
to simulate than to solve a partial differential equation as the
Fokker-Planck equation.

This paper is notationally a bit cumbersome. This is quite an
inescapable situation given that we are dealing with five continuous
hydrodynamic fields. In order to alleviate a bit this we have decided
to give the structure of the Fokker-Planck equation and the
corresponding stochastic differential equation for arbitrary fields
before particularizing to the case of hydrodynamic fields. In any
case, the final results have a very compact structure closely
reminiscent to the Landau and Lifshitz theory.

The paper is distributed as follows. In section II we discuss the
definition of the coarse-grained hydrodynamic variables in some
detail. In section III we rewrite the standard FPE for the case that
the set of relevant variables are extensive variables characterized by
a discrete index associated to a lattice (i.e. a field defined on a
lattice). We also write the FPE equation corresponding to the
densities of the extensive variables. This FPE is then expressed in
terms of functional derivatives. Section IV is devoted to the
equivalent SDE for fields on a lattice and its corresponding
functional Langevin equation where precise meaning to the white noise
term is given.  In section V, the general FPE for fields on a lattice
obtained in section III is particularized to the case that the fields
are the hydrodynamic fields. Section VI is devoted to the derivation
of the SDE for hydrodynamic fields. Finally, we discuss the results in
the last section.

\section{Hydrodynamic variables}
Our aim is to obtain the FPE for the hydrodynamic continuous
fields. From a mathematical point of view there are subtleties
regarding the existence of the continuum limit for stochastic
variables and for this reason we will restrict ourselves to the case
that the fields are defined on a lattice. In this section we discuss
the discrete variables that are representative of the hydrodynamic
fields. A word is in order about the definition of the lattice of
points ${\bf r}$. This lattice is usually taken as a regular cubic
lattice \cite{saa82}. In the formulation that follows there is no
restriction, however, to the kind of lattice that can be used and
irregular lattices are also possible. To each lattice point we
associate a {\em volume element} of volume $v_{\bf r}$ that defines
the region of space associated to each lattice point. In the case of
the cubic lattice it is natural to associate a cubic cell of volume
$\epsilon^3$. For irregular lattices, the Voronoi tessellation allows
to define non-overlapping volume elements that cover all the space
\cite{esp97}. Note that the cubic cell is a special case of Voronoi
tessellation for the case of a regular cubic lattice.

We take as relevant variables the total mass $M_{\bf r}$, the total
momentum ${\bf P}_{\bf r}$ and total energy $E_{\bf r}$ of that set of
particles that are within the volume element associated to the lattice
point ${\bf r}$. In compact form, $M_{\bf r},{\bf P}_{\bf r},E_{\bf
r}\rightarrow A_{\bf r}^m$ with $m=0,\ldots,4$. Because these
variables are conserved variables, one expects that their dynamics is
much slower than any other set of dynamic variables in such a way that
they form a closed description for the system. The size of the volume
elements should be so large that the relevant variables can be
considered as continuous. For example, if a single particle enters the
cell, the variation of the mass of the cell can be considered as
infinitesimal.

In mathematical terms, the relevant variables are given by

\begin{eqnarray}
M_{\bf r}&=& \sum_i m\chi_{\bf r}({\bf q}_i)
\nonumber\\
{\bf P}_{\bf r}&=&\sum_i{\bf p}_i\chi_{\bf r}({\bf q}_i)
\nonumber\\
E_{\bf r}&=&\sum_ie_i\chi_{\bf r}({\bf q}_i)
\label{e1c}
\end{eqnarray}
where we denote by ${\bf q}_i,{\bf p}_i$ the position and momentum of
particle $i$ and its energy is
\begin{equation}
e_i=\frac{{\bf p}_i^2}{2m}
+\frac{1}{2}\sum^{'}_j\phi({\bf q}_i,{\bf q}_j)
\label{e1b}
\end{equation}
The function $\chi_{\bf r}({\bf q})$ is the characteristic function of
the cell centered at ${\bf r}$, which takes values $1$ if ${\bf q}$
belongs to the cell and $0$ otherwise. The volume of the cell at ${\bf
r}$ is given by

\begin{equation}
v_{\bf r} = \int d{\bf r}'\chi_{\bf r}({\bf r}')
\label{vol}
\end{equation}
If we displace the position ${\bf q}_i$ of all the particles a vector
${\bf r}'$ and simultaneously displace the location of the cell at
${\bf r}$ the same vector ${\bf r}'$, the numerical value of
$\chi_{\bf r}({\bf q}_i)$ will not change. This translational
invariance implies that $\chi_{\bf r}({\bf q}_i)=\chi_{\bf r}({\bf
q}_i-{\bf r})$. It is apparent that if all cells are identical in
shape as occurs in regular lattices we can suppress the label ${\bf r}$
of the characteristic function, i.e.  $\chi_{\bf r}({\bf q}_i-{\bf
r})=\chi({\bf q}_i-{\bf r})$.

The density variables associated to our selected extensive variables
are

\begin{eqnarray}
\rho_{\bf r}&=&\frac{M_{\bf r}}{v_{\bf r}}
\equiv\sum_i m\overline{\delta}_{\bf r}({\bf r}-{\bf q}_i)
\nonumber\\
{\bf g}_{\bf r}&=&\frac{{\bf P}_{\bf r}}{v_{\bf r}}
\equiv \sum_i{\bf p}_i\overline{\delta}_{\bf r}({\bf r}-{\bf q}_i)
\nonumber\\
e_{\bf r}&=&\frac{E_{\bf r}}{v_{\bf r}}
\equiv \sum_ie_i\overline{\delta}_{\bf r}({\bf r}-{\bf q}_i)
\label{dens}
\end{eqnarray}
where we have introduced 

\begin{equation}
\overline{\delta}_{\bf r}({\bf
r}')=\chi_{\bf r}({\bf r}')/v_{\bf r}
\label{intro}
\end{equation}
that satisfies by virtue of (\ref{vol}) the normalization condition

\begin{equation}
\int d{\bf r}'\overline{\delta}_{\bf r}({\bf r}')=1
\label{norm}
\end{equation}
The velocity field is defined by

\begin{equation}
{\bf v}_{\bf r}\equiv\frac{{\bf g}_{\bf r}}{\rho_{\bf r}}
\label{e6}
\end{equation}
Note that $\sum_{\bf r} v_{\bf r} \rho_{\bf r} = \sum_{\bf r} M_{\bf
r}$, etc. We want to have $\sum_{\bf r} M_{\bf r}$ equal to the total
mass of the systems and this is only possible if the cells are
non-overlapping and cover all the space. If this condition is not met,
then $\sum_{\bf r} M_{\bf r}$ is not a conserved variable and one does
not expect it to be a slow variable.

For the case of a cubic lattice with cubic volume elements, the
characteristic function can be represented as a product of square step
functions in the $x,y,z$ directions. One can argue that if the
particles have some finite size, then it is meaningful to smooth out
the step functions at the borders of the cell and to talk about
fractions of a particle being in different cells. Actually, one could
use coarsening functions other than $\overline{\delta}_{\bf r}({\bf
r}')$ or these smoothed square step functions defining the ``shape''
of the cell. Zubarev and Morozov make use of a particular
coarse-graining procedure in which the hydrodynamic fields are
represented in a Fourier basis and then only the small $k<k_0$
components are retained \cite{zub83}. In real space this is equivalent
to use a coarse graining function of the form

\begin{equation}
\chi(r)
=\frac{k_0^3}{2\pi^2}
\left[\frac{\sin k_0r}{(k_0r)^3}-\frac{\cos k_0r}{(k_0r)^2}\right]
\label{e2}
\end{equation}
This particular coarse-graining produces a coarse-grained density
field that is not everywhere positive. This can be cured, for example,
by using a Gaussian function

\begin{equation}
\Delta(r)=\frac{1}{\pi^{3/2}}\exp-\left(r/\epsilon\right)^2
\label{e3}
\end{equation}
where $\epsilon$ is a coarse-graining length. Nevertheless, in this
paper we will restrict ourselves to the use of the coarse-graining
function $\overline{\delta}_{\bf r}({\bf r'})$ derived from the
characteristic function of each cell, Eqn. (\ref{intro}).

We will need the time derivatives of the hydrodynamic variables. They
have the form
\begin{eqnarray}
iL\rho_{\bf r} &=& -\nabla_{\bf r}\!\cdot\!{\bf g}_{\bf r}
\nonumber\\
iL{\bf g}_{\bf r} &=& -\nabla_{\bf r}\!\cdot\!\sigma_{\bf r}
\nonumber\\
iL   e_{\bf r} &=& -\nabla_{\bf r}\!\cdot\!\tau_{\bf r}
\label{e9b}
\end{eqnarray}
where the coarse-grained stress tensor and energy flux are given by
the usual expressions (see, for example, Ref. \cite{gra82}).

A remark on the meaning of $\nabla_{\bf r}$ is in order. In obtaining
(\ref{e9b}) we have used
\begin{equation}
\frac{\partial}{\partial {\bf q}_i}\overline{\delta}_{\bf r}
({\bf r}-{\bf q}_i)=
-\frac{\partial}{\partial {\bf r}}\overline{\delta}_{\bf r}
({\bf r}-{\bf q}_i)
\label{nabla}
\end{equation}
where we assume that the index ${\bf r}$ is continuous. In fact,
Eqn. (\ref{nabla}) is a mathematical identity. However, we will
evaluate the right hand side of (\ref{nabla}) only at the discrete
values of the vector ${\bf r}$. There is no need to use discretized
forms for the nabla operator with this understanding \cite{saa82}.

\section{FPE for fields}
Zwanzig derived from first principles the Fokker-Planck equation for
the distribution function of a discrete set of variables with the aid
of a projection operator technique \cite{zwa61}. The derivation
provided explicit expressions for the drift and diffusion terms of the
equation in terms of microscopic quantities. In this section, we will
present the FPE for the case that the set of discrete variables are
extensive variables $A^m_{\bf r}$ (or simply $A$) defined in each
cell. We will also give the FPE for the distribution function of the
{\em densities} $\tilde{A}^m_{\bf r}=A^m_{\bf r}/v_{\bf r}$.

Let us consider an isolated system governed by Hamilton's equation of
motion ($H(z)$ is the Hamiltonian and $iL$ is Liouville's operator for
the dynamics). The microscopic degrees of freedom are denoted by
$z$. The FPE is a differential equation for the probability density
$p(\alpha,t)$ that the system takes values $\alpha$ at time $t$ for
the phase functions $A(z)$, {\em i.e.}
\begin{equation}
p(\alpha,t) \equiv\int\rho_t(z) \delta(A(z)-\alpha)dz
\label{ch22}
\end{equation}
where $\delta(A(z)-\alpha)$ is a product of Dirac's delta functions,
one for each relevant variable, and $\rho_t(z)$ is the solution of
Liouville's equation. 

The FPE is \cite{zub83,esp93,zwa61,gra82}
\begin{eqnarray}
\partial_t p(\alpha,t)
&=&-\sum_{m{\bf r}}\frac{\partial}{\partial \alpha_{\bf r}^m}
K_{\bf r}^m(\alpha)p(\alpha,t)
\nonumber\\
&+&
\frac{1}{2}\sum_{m{\bf r}}\sum_{m'{\bf r}'}
\frac{\partial}{\partial \alpha_{\bf r}^m}
 \frac{\partial}{\partial \alpha_{{\bf r}'}^{m'}}
D_{{\bf r}{\bf r}'}^{mm'}(\alpha)p(\alpha,t)
\label{fp1}
\end{eqnarray}
with the following definitions 

\begin{eqnarray}
K_{\bf r}^m(\alpha)&\equiv& v_{\bf r}^m(\alpha)
+\sum_{m'{\bf r}'}\zeta_{{\bf r}{\bf r}'}^{mm'}(\alpha)
F_{{\bf r}'}^{m'}(\alpha)
\nonumber\\
&+&\sum_{m'{\bf r}'}\frac{\partial}{\partial\alpha_{\bf r}^m}
\zeta_{{\bf r}{\bf r}'}^{mm'}(\alpha)
\label{ch2driftb}
\end{eqnarray}
with
\begin{eqnarray}
v_{\bf r}^m(\alpha)&\equiv&\langle iLA^m_{\bf r}\rangle^{\alpha}
\nonumber\\
\zeta_{{\bf r}{\bf r}'}^{mm'}(\alpha)&\equiv&
\int_0^{\infty}du
\langle (\delta iLA_{\bf r}^m
\delta iLA_{{\bf r}'}^{m'}(u)\rangle^{\alpha}
\nonumber\\
F_{\bf r}^m(\alpha)&\equiv&
\frac{\partial}{\partial\alpha_{\bf r}^m}\ln \Omega(\alpha)
\nonumber\\
\Omega(\alpha)&\equiv&\int \delta(A(z)-\alpha)dz 
\nonumber\\
D_{{\bf r}{\bf r}'}^{mm'}(\alpha)&\equiv& 
\zeta_{{\bf r}{\bf r}'}^{mm'}(\alpha)
+\zeta_{{\bf r}'{\bf r}}^{m'm}(\alpha)
\label{ch210b}
\end{eqnarray}
where we have defined the projected part
$\delta iLA^m_{\bf r}\equiv iLA^m_{\bf r}
-\langle iLA^m_{\bf r}\rangle^{\alpha}$. 

We have introduced the average of an arbitrary phase function $G$
over the submanifold $A(z)=\alpha$ by
\begin{equation}
\langle G\rangle^{\alpha}\equiv\frac{1}{\Omega(\alpha)}
\int \delta(A(z)-\alpha)G(z)dz 
\label{ch25}
\end{equation}
The derivation of the FPE from the underlying microscopic dynamics of
the system relies on the fact that the relevant variables $A$ are
much slower than the rest of variables of the system. This allows for
a Markovian memoryless description of the dynamics of these relevant
variables. A systematic expansion of a microscopically exact master
equation in terms of a slowness parameter (which is, essentially, the
ratio between time scales between relevant and irrelevant variables)
has been carried out by Mori et al. \cite{mor74}. To second order
in the slowness parameter the non-linear FPE (\ref{fp1}) is obtained.

In the microscopic formulation of the FPE the drift $K^m_{\bf r}$ and
the diffusion tensor $D^{mm'}_{{\bf r}{\bf r'}}$ are defined in terms
of microscopic quantities. The problem is that these quantities are
difficult to compute. Basically, three difficulties have to be
resolved: the calculation of the averages
$\langle\cdots\rangle^\alpha$, the calculation of the thermodynamic
forces $F^m_{\bf r}$ (or equivalently of the phase space volume
$\Omega(\alpha)$) and finally the calculation of the correlation
functions in the kinetic tensor $\zeta^{mm'}_{{\bf r}{\bf r'}}$. The
first two are closely related and can be resolved for the case of
hydrodynamic variables by resorting to the local equilibrium
approximation. The explicit calculation of the kinetic coefficients is
much more difficult and they are simply left as phenomenological
coefficients.

The rest of the section is devoted to the FPE equation for the
distribution function of the densities $\tilde{A}^m_{\bf r}=A^m_{\bf
r}/v_{\bf r}$. The corresponding FPE is obtained by a simple change of
variables

\begin{eqnarray}
\partial_t p(v_{\bf r}\tilde{\alpha},t)&=&-\sum_{m{\bf r}}
\frac{\partial}{\partial\tilde{\alpha}_{\bf r}^m}\frac{1}{v_{\bf r}} 
K_{\bf r}^m(v_{\bf r}\tilde{\alpha})p(v_{\bf r}\tilde{\alpha},t)
\nonumber\\
&+&
\frac{1}{2}\sum_{m{\bf r}}\sum_{m'{\bf r}'}
\frac{\partial}{\partial\tilde{\alpha}_{\bf r}^m}
\frac{\partial}{\partial\tilde{\alpha}_{{\bf r}'}^{m'}}
\frac{1}{v_{\bf r}^2} D_{{\bf r}{\bf r}'}^{mm'}
(v_{\bf r}\tilde{\alpha})p(v_{\bf r}\tilde{\alpha},t)
\nonumber\\
\label{fp3}
\end{eqnarray}
where $\tilde{\alpha}^m_{\bf r}=\alpha^m_{\bf r}/v_{\bf r}$. We
therefore introduce in a natural way the following quantities
\begin{eqnarray}
P[\tilde{\alpha},t]&\equiv&\prod^M_{\bf r} v_{\bf r} 
p(v\tilde{\alpha},t)
\nonumber\\
K_{\bf r}^m[\tilde{\alpha}]&\equiv&\frac{1}{v_{\bf r}}
K_{\bf r}^m(v\tilde{\alpha})
\nonumber\\
D_{{\bf r}{\bf r}'}^{mm'}[\tilde{\alpha}]&=&\frac{1}{v_{\bf r}^2}
D_{{\bf r}{\bf r}'}^{mm'}(v\tilde{\alpha})
\nonumber\\
F^m_{\bf r}[\tilde{\alpha}]&\equiv& F^m_{\bf r}(v\tilde{\alpha})
\nonumber\\
v^m_{\bf r}[\tilde{\alpha}]&\equiv&\frac{1}{v_{\bf r}}v^m_{\bf r}
(v\tilde{\alpha})
\nonumber\\
\zeta^{mm'}_{{\bf r}{\bf r}'}[\tilde{\alpha}]&\equiv&
\frac{1}{v_{\bf r}^2}
\zeta^{mm'}_{{\bf r}{\bf r}'}(v\tilde{\alpha})
\nonumber\\
\Omega[\tilde{\alpha}]&\equiv&\Omega(v\tilde{\alpha})
\label{fp5}
\end{eqnarray}
We use square brackets to denote that the left hand sides have
different functional forms from the right hand side (for example,
$F^m_{\bf r}[\tilde{\alpha}]$ is a different function of its argument
from $F^m_{\bf r}(v\tilde{\alpha})$ even though they are named with
the same symbol $F^m_{\bf r}$). $M$ is the total number of
variables. In this way the probability density $P[\tilde{\alpha},t]$
is normalized to unity, {\em i.e.}, $\int {\cal D}\tilde{\alpha}
P[\tilde{\alpha},t]=1$, where ${\cal D} \tilde{\alpha}$ is a short
hand for $d\tilde{\alpha}^M$.

Note that these definitions (\ref{fp5}) allow to express the
reversible part of the drift and the kinetic tensor in terms of
averages over density fields
\begin{eqnarray}
v^m_{\bf r}[\tilde{\alpha}]
&=&\langle iL\tilde{A}^m_{\bf r}\rangle^{v\tilde{\alpha}}
\nonumber\\
\zeta^{mm'}_{{\bf r}{\bf r}'}[\tilde{\alpha}]
&=&\int_0^\infty \!\!du\;\;
\langle
(\delta iL\tilde{A}^m_{\bf r})
(\delta iL\tilde{A}^{m'}_{{\bf r}'}(u))\rangle^{v\tilde{\alpha}}
\label{fp7}
\end{eqnarray}
where we have defined the projected part
$\delta iL\tilde{A}^m_{\bf r}\equiv iL\tilde{A}^m_{\bf r}-\langle
iL\tilde{A}^m_{\bf r}\rangle^{v\tilde{\alpha}}$. 
With the definitions (\ref{fp5}) the FPE  for the density
fields is

\begin{eqnarray}
\partial_t P[\tilde{\alpha},t]&=&-\sum_{m{\bf r}}
\frac{\partial}{\partial\tilde{\alpha}_{\bf r}^m}
K_{\bf r}^m[\tilde{\alpha}]P[\tilde{\alpha},t]
\nonumber \\
&+&
\frac{1}{2}\sum_{m{\bf r}}\sum_{m'{\bf r}'}
\frac{\partial}{\partial\tilde{\alpha}_{\bf r}^m}
\frac{\partial}{\partial\tilde{\alpha}_{{\bf r}'}^{m'}}
D_{{\bf r}{\bf r}'}^{mm'}[\tilde{\alpha}]P[\tilde{\alpha},t]
\label{fpdens}
\end{eqnarray}
with 
\begin{eqnarray}
K_{\bf r}^m[\tilde{\alpha}]&=&v^m_{\bf r}[\tilde{\alpha}]
+\sum_{{\bf r}'}v_{\bf r}\zeta^{mm'}_{{\bf r}{\bf r}'}
[\tilde{\alpha}]F^{m'}_{{\bf r}'}[\tilde{\alpha}]
\nonumber\\
&+&\sum_{{\bf r}'}
\frac{\partial}{\partial \tilde{\alpha}^{m'}_{{\bf r}'}}
\zeta^{mm'}_{{\bf r}{\bf r}'}[\tilde{\alpha}]
\nonumber\\
F^m_{{\bf r}'}[\tilde{\alpha}]&=&\frac{1}{v_{\bf r}}
\frac{\partial}{\partial \tilde{\alpha}^m_{{\bf r}'}}
\ln\Omega[\tilde{\alpha}]
\nonumber\\
D_{{\bf r}{\bf r}'}^{mm'}[\tilde{\alpha}]
&=&\zeta^{mm'}_{{\bf r}{\bf r}'}[\tilde{\alpha}]
+\zeta^{m'm}_{{\bf r}'{\bf r}}[\tilde{\alpha}]
\label{def}
\end{eqnarray}
The FPE for the densities (\ref{fpdens}) admits a suggestive continuum
notation. By making the substitutions of partial derivatives by
functional derivatives according to
\begin{equation}
\frac{1}{v_{\bf r}}
\frac{\partial}{\partial\tilde{\alpha}_{{\bf r}}^{m}}
\rightarrow\frac{\delta}{\delta\tilde{\alpha}_{\bf r}^m}
\end{equation}
and sums by integrals
\begin{equation}
\sum_{\bf r} v_{\bf r}\rightarrow \int\!\! d{\bf r}
\label{subs}
\end{equation}
we can write equation (\ref{fpdens}) in functional form

\begin{eqnarray}
\partial_t P[\tilde{\alpha},t]
&=&-\int\!\! d{\bf r}\frac{\delta}{\delta \tilde{\alpha}_{\bf r}^m}
K_{\bf r}^m[\tilde{\alpha}]P[\tilde{\alpha},t]
\nonumber\\
&+&
\frac{1}{2}\int\!\! d{\bf r}\int\!\! d{\bf r}'
\frac{\delta}{\delta\tilde{\alpha}_{\bf r}^m}
\frac{\delta}{\delta\tilde{\alpha}_{{\bf r}'}^{m'}}
D_{{\bf r}{\bf r}'}^{mm'}[\tilde{\alpha}]P[\tilde{\alpha},t]
\label{fp4}
\end{eqnarray}
with
\begin{eqnarray}
K_{\bf r}^m[\tilde{\alpha}]&=&v^m_{\bf r}[\tilde{\alpha}]
+\int d{\bf r}'\zeta^{mm'}_{{\bf r}{\bf r}'}[\tilde{\alpha}]
F^{m'}_{{\bf r}'}[\tilde{\alpha}]
\nonumber\\
&+&\int d{\bf r}'\frac{\delta}{\delta \tilde{\alpha}^{m'}_{{\bf r}'}}
\zeta^{mm'}_{{\bf r}{\bf r}'}[\tilde{\alpha}]
\nonumber\\
F^m_{{\bf r}'}[\tilde{\alpha}]&=&
\frac{\delta}{\delta \tilde{\alpha}^m_{{\bf r}'}}
\ln\Omega[\tilde{\alpha}]
\nonumber\\
D_{{\bf r}{\bf r}'}^{mm'}[\tilde{\alpha}]
&=&\zeta^{mm'}_{{\bf r}{\bf r}'}[\tilde{\alpha}]
+\zeta^{m'm}_{{\bf r}'{\bf r}}[\tilde{\alpha}]
\label{deffunc}
\end{eqnarray}
The functional FPE (\ref{fp4}) is understood in this work simply as a
notational device and the mathematically well-defined FPE is the
discrete FPE (\ref{fpdens}).

\section{Stochastic Differential Equation for fields}
As it is well known \cite{gar83}, associated to the FPE there is a
stochastic differential equation (SDE) describing the stochastic
process $\alpha(t)$. The SDE corresponding to the FPE (\ref{fpdens})
is given by

\begin{equation}
d\tilde{\alpha}^m_{\bf r}(t) = K^m_{\bf r}[\tilde{\alpha}(t)] dt 
+ \sum_{m'{\bf r}'}B^{mm'}_{{\bf r}{\bf r}'}
[\tilde{\alpha}(t)]dW^{m'}_{{\bf r}'}(t)
\label{cont6}
\end{equation}
where we have introduced the independent increments of the Wienner
process associated to each cell ${\bf r}$. These increments satisfy
the mnemotechnical rule for It\^o calculus

\begin{equation}
dW^m_{\bf r}(t)dW^{m'}_{{\bf r}'}(t')=\delta_{mm'}\delta_{{\bf r}{\bf
r}'}dt
\quad\quad\quad \mbox{if $t=t'$} 
\label{cont7}
\end{equation}
and zero if $t\neq t'$. The Kronecker delta $\delta_{{\bf r}{\bf r}'}$
is one if ${\bf r}={\bf r'}$ and zero otherwise. The matrix
$B^{mm'}_{{\bf r}{\bf r}'}[\tilde{\alpha}]$ is the square root in
matrix sense of $D^{mm'}_{{\bf r}{\bf r}'}[\tilde{\alpha}]$, this is

\begin{equation}
\sum_{m'{\bf r}'}B^{mm'}_{{\bf r}{\bf r}'}
[\tilde{\alpha}]B^{m''m'}_{{\bf r}''{\bf r}'}[\tilde{\alpha}]
=D^{mm''}_{{\bf r}{\bf r}''}[\tilde{\alpha}]
\label{sqr}
\end{equation}

The SDE (\ref{cont6}) is
interpreted in It\^o sense. This interpretation is most convenient for
numerical simulations because the increment of the Wienner process is
uncorrelated with the variables at the same time. For a discussion of
the advantages of deriving stochastic differential equations from
first principles through the FPE route instead of directly from the
microscopic dynamics, see Ref. \cite{esp93}.

A physically suggestive way of writing the SDE (\ref{cont6}) is by
using white noise in space and time defined as

\begin{equation}
\eta^m({\bf r},t)\equiv \frac {dW^m_{\bf r}(t)}{\sqrt{v_{\bf r}}dt}
\label{white}
\end{equation}
We thus write Eqn. (\ref{cont6}) as an ordinary differential equation, 
i.e. in the form of a Langevin equation
\begin{equation}
\partial_t\tilde{\alpha}^n_{\bf r}(t) = K^m_{\bf r}[\tilde{\alpha}(t)] 
+ \int\!\! d{\bf r}'\tilde{B}^{mm'}_{{\bf r}{\bf r}'}
[\tilde{\alpha}(t)]\eta^{m'}({\bf r}',t)
\label{cont8}
\end{equation}
where we have introduced the ``continuous'' matrix 
$\tilde{B}^{mm'}_{{\bf r}{\bf r}'}[\alpha(t)]
=B^{mm'}_{{\bf r}{\bf r}'}[\alpha(t)]/\sqrt{v_{\bf r}}$
that satisfies

\begin{equation}
\int\!\! d{\bf r}'\tilde{B}^{mm'}_{{\bf r}{\bf r}'}
[\tilde{\alpha}]\tilde{B}^{m''m'}_{{\bf r}''{\bf r}'}[\tilde{\alpha}]
=D^{mm''}_{{\bf r}{\bf r}''}[\tilde{\alpha}]
\label{sqrcont}
\end{equation}

The white noise term has the following correlation

\begin{eqnarray}
\langle \eta^m({\bf r},t)\eta^{m'}({\bf r}',t')\rangle &=&
\langle \frac {dW^m_{\bf r}(t)}{\sqrt{v_{\bf r}}dt}
        \frac {dW^{m'}_{{\bf r}'}(t')}{\sqrt{v_{\bf r}}dt'}\rangle
\nonumber\\
&=&\delta_{mm'}\frac{\delta_{{\bf r}{\bf r}'}}{v_{\bf r}}\frac{1}{dt}
\quad\quad\quad \mbox{if $t=t'$} 
\label{corr1}
\end{eqnarray}
and zero if $t\neq t'$. We can rewrite this correlation as follows

\begin{equation}
\langle \eta^m({\bf r},t)\eta^{m'}({\bf r}',t')\rangle 
=\delta_{mm'}\delta({\bf r}-{\bf r}')\delta(t-t')
\label{corr}
\end{equation}
The reason for such a procedure is that expression (\ref{corr}) will
give the same results as (\ref{corr1}) when computing averages and
correlations. Indeed, let us consider the noise term in
(\ref{cont6}) which  is given by

\begin{equation}
f^m_{\bf r}(t)dt \equiv \sum_{{\bf r}'}B^{mm'}_{{\bf r}{\bf r}'}
[\alpha(t)]dW^{m'}({\bf r},t)
\label{noisdisc}
\end{equation}
The notation suggest that $f^m_{\bf r}(t)$ is related to the
derivative of the Wienner process that does not exist in a strict
sense. Nevertheless, with the understanding that $f^m_{\bf r}(t)dt$
will appear always inside integral expressions (which will be
interpreted as stochastic integrals) such a procedure leads to correct
results. For example, the correlation of the noise terms
(\ref{noisdisc}) will be

\begin{eqnarray}
&&\left\langle\int\!\!dt\int\!\!dt''f^m_{\bf r}(t)
 f^{m''}_{{\bf r}''}(t'')\right\rangle
\nonumber\\
&=&\left\langle\int\!\!\int
\sum_{{\bf r}'}B^{mm'}_{{\bf r}{\bf r}'}
[\tilde{\alpha}(t)]dW^{m'}({\bf r}',t)\right.
\nonumber\\
&\times&\left.\sum_{{\bf r}'''}B^{m''m'''}_{{\bf r}''{\bf r}'''}
[\tilde{\alpha}(t')]dW^{m'''}({\bf r}''',t')\right\rangle
\nonumber\\
&=&\left\langle\int\!\!dt
\sum_{{\bf r}'}B^{mm'}_{{\bf r}{\bf r}'}[\tilde{\alpha}(t)]
B^{m''m'}_{{\bf r}''{\bf r}'}[\tilde{\alpha}(t
)]\right\rangle
\nonumber\\
&=& \int dt \int {\cal D}\tilde{\alpha} P[\tilde{\alpha},t] 
D^{mm''}_{{\bf r}{\bf r}''}[\tilde{\alpha}]
=\int dt \left\langle D^{mm''}_{{\bf r}{\bf r}''}\right\rangle_t
\label{know}
\end{eqnarray}
where It\^o rule (\ref{cont7}) has been used. Note that one can write

\begin{equation}
\left\langle f^m_{\bf r}(t) f^{m''}_{{\bf r}''}(t'')\right\rangle
=\delta(t-t'')\left\langle D^{mm''}_{{\bf r}{\bf r}''}\right\rangle_t
\label{know1}
\end{equation}
because if one integrates over $t,t''$ one recovers (\ref{know}) and
the integration limits are arbitrary.

On the other hand, in the continuous version (\ref{cont8}) the noise
term is given by
\begin{equation}
f^m_{\bf r}(t) \equiv 
\int d{\bf r}'\tilde{B}^{mm'}_{{\bf r}{\bf r}'}
[\tilde{\alpha}(t)]\eta^{m'}({\bf r}',t)
\label{noiscont}
\end{equation}
Now, if we use the continuum version (\ref{noiscont}) and proceed with
the ordinary calculus rules (and the fact that the noise is
uncorrelated with the variables at the same time) we will have
\begin{eqnarray}
&&\left\langle \int\!\!dt\int\!\!dt'f^m_{\bf r}(t) 
f^{m''}_{{\bf r}''}(t')\right\rangle
\nonumber\\
&=&\left\langle\int\!\!dt\int\!\!dt'\int d{\bf r}'
\tilde{B}^{mm'}_{{\bf r}{\bf r}'}[\tilde{\alpha}(t)]
\eta^{m'}({\bf r}',t)\right.
\nonumber\\
&\times&\left.\int d{\bf r}'''\tilde{B}^{m''m'''}_{{\bf r}''{\bf r}'''}
[\tilde{\alpha}(t')]\eta^{m'''}({\bf r}''',t')\right\rangle
\nonumber\\
&=&\int\!\!dt\int\!\!dt'\int d{\bf r}'\int d{\bf r}'''
\left\langle \tilde{B}^{mm'}_{{\bf r}{\bf r}'}[\tilde{\alpha}(t)]
\tilde{B}^{m''m'''}_{{\bf r}''{\bf r}'''}[\tilde{\alpha}(t')]
\right\rangle
\nonumber\\
&\times&\left\langle \eta^{m'}({\bf r}',t)
\eta^{m'''}({\bf r}''',t')\right\rangle
\nonumber\\
&=&\int\!\!dt\int d{\bf r}'
\left\langle \tilde{B}^{mm'}_{{\bf r}{\bf r}'}[\tilde{\alpha}(t)]
\tilde{B}^{m''m'}_{{\bf r}''{\bf r}'}[\tilde{\alpha}(t)]\right\rangle
\nonumber
\\
&=& \int dt\left\langle  D^{mm'}_{{\bf r}{\bf r}'}
[\tilde{\alpha}(t)]\right\rangle
\label{ok}
\end{eqnarray}
which provides the same result as (\ref{know}).

\section{Fokker-Planck equation for a simple fluid}
Now we wish to specify the FPE (\ref{fpdens}) for the case that the
continuum variables $\tilde{A}^m_{\bf r}$ are the hydrodynamic
variables.  This has been done explicitly in Ref.\cite{zub83} for the
case that the coarse-grained hydrodynamic fields are defined in terms
of the small $k<k_0$ Fourier components. In order to derive the SDE
(\ref{cont8}) we found more convenient to work with the coarse-grained
hydrodynamic fields defined in terms of the densities (i.e. extensive
variables divided by the volume of the cell). For this case the
derivation of the explicit forms for the drift and diffusion terms
follows very similar lines as those of Ref. \cite{zub83} and
therefore, only the final results are quoted here. The physical
assumptions underlying the derivation of the Fokker-Planck equation
for hydrodynamic variables  are essentially that the fields are 
slowly varying in space and time. Therefore, a second order expansion in
gradients is allowed along with a Markovian assumption. Consistently,
the concept of local equilibrium is advocated. These assumptions are
thoroughly discussed by Zubarev and Morozov \cite{zub83}.

The thermodynamic forces $F^m_{\bf r}[\tilde{\alpha}]$ are given by

\begin{eqnarray}
F^0_{\bf r}[\tilde{\alpha}]
&=&-\frac{1}{T_{\bf r}}\left(\mu_{\bf r}
-\frac{{\bf v}_{\bf r}^2}{2}\right)
\nonumber\\
F^\nu_{\bf r}[\tilde{\alpha}]
&=&-\frac{{\bf v}^\nu_{\bf r}}{T_{\bf r}}
\nonumber\\
F^4_{\bf r}[\tilde{\alpha}]
&=&\frac{1}{T_{\bf r}}\equiv \beta_{\bf r}
\label{intens0}
\end{eqnarray}
where $T_{\bf r}$ is the temperature and $\mu_{\bf r}$ is the chemical
potential which are given in terms of the equilibrium equations
of state, this is

\begin{eqnarray}
T_{\bf r}&=& T^{eq}(\rho_{\bf r},\epsilon_{\bf r})
\nonumber\\
\mu_{\bf r}&=& \mu^{eq}(\rho_{\bf r},\epsilon_{\bf r})
\label{eqstate}
\end{eqnarray}
where the internal energy is $\epsilon\equiv e-\frac{1}{2}\rho{\bf
v}^2$.  Note that the thermodynamic parameters $F^m_{\bf
r}[\tilde{\alpha}]$ which are, in principle, functionals of the
fields become local functions of the density variables, this is,
$F^m_{\bf r}[\tilde{\alpha}]=F^m(\tilde{\alpha}_{\bf r})$.

The reversible part $v^m_{\bf r}[\tilde{\alpha}]$ of
the drift term (\ref{fp7}) is given by

\begin{eqnarray}
\langle iL\rho_{\bf r}\rangle^{\tilde{\alpha}}
&=&-\nabla\!\cdot\!{\bf g}_{\bf r}
\nonumber\\
\langle iL{\bf g}_{\bf r}\rangle^{\tilde{\alpha}}
&=&-\nabla\!\cdot\!\left[{\bf v}_{\bf r}{\bf v}_{\bf r}\rho_{\bf r}
+p_{\bf r}{\bf 1}\right]
\nonumber\\
\langle iLe_{\bf r}\rangle^{\tilde{\alpha}}
&=&-\nabla\!\cdot\!\left[{\bf v}_{\bf r}(e_{\bf r}+p_{\bf r})\right]
\label{leres}
\end{eqnarray}
where $p_{\bf r}$ is the equilibrium pressure evaluated at the
instantaneous values of the hydrodynamic fields, this is $p_{\bf
r}=p^{eq}(\rho_{\bf r},\epsilon_{\bf r})$.

The kinetic tensor is given by (\ref{fp7}). Note that the components
corresponding to the density field vanish because this field has no
projected part, i.e. $\langle iL\rho_{\bf
r}\rangle^{\tilde{\alpha}}=iL\rho_{\bf r}$, see (\ref{leres}). This
imply that the only components different from zero correspond to the
momentum density and the energy density. The explicit forms are

\begin{eqnarray}
\zeta^{\alpha\beta}_{{\bf r}{\bf r}'}&=&
\partial_{{\bf r}_\mu}\partial_{{\bf r}'_\nu}
\delta({\bf r}-{\bf r'})L^{\alpha\mu\beta\nu}_{\bf r}
\nonumber\\
\zeta^{\alpha 4 }_{{\bf r}{\bf r}'}&=&
\partial_{{\bf r}_\mu}\partial_{{\bf r}'_\nu}
\delta({\bf r}-{\bf r'})
L^{\alpha\mu\beta\nu}_{\bf r}{\bf v}^{\beta}_{\bf r}
\nonumber\\
\zeta^{4\alpha }_{{\bf r}{\bf r}'}&=&
\partial_{{\bf r}_\mu}\partial_{{\bf r}'_\nu}
\delta({\bf r}-{\bf r'})
L^{\beta\mu\alpha\nu}_{\bf r}{\bf v}^{\beta}_{\bf r}
\nonumber\\
\zeta^{44}_{{\bf r}{\bf r}'}&=&
\partial_{{\bf r}_\mu}\partial_{{\bf r}'_\nu}
\delta({\bf r}-{\bf r'})
\left[L^{\mu\nu}_{\bf r}+L^{\alpha\mu\beta\nu}_{\bf r}
{\bf v}^{\alpha}_{\bf r}{\bf v}^{\beta}_{\bf r}\right]
\nonumber\\
\label{q2b}
\end{eqnarray}
where we have  introduced
\begin{eqnarray}
L^{\mu\nu}_{\bf r}&\equiv& T({\bf r})\kappa({\bf r})\delta_{\mu\nu}
\nonumber\\
L^{\alpha\beta\mu\nu}_{\bf r}
&\equiv&
 T({\bf r})\eta({\bf r}) g_{\alpha\beta\mu\nu}
+T({\bf r})\zeta({\bf r})\delta_{\alpha\beta\mu\nu}
\nonumber\\
g_{\alpha\beta\mu\nu}&\equiv&\delta_{\alpha\mu}\delta_{\beta\nu}
+\delta_{\alpha\nu}\delta_{\beta\mu}
-\frac{2}{3}\delta_{\alpha\beta}\delta_{\mu\nu}
\nonumber\\
\delta_{\alpha\beta\mu\nu}&\equiv&\delta_{\alpha\beta}\delta_{\mu\nu}
\label{ten4}
\end{eqnarray}
where $\eta({\bf r}),\zeta({\bf r})$ are the shear and bulk
viscosities and $\kappa({\bf r})$ is the thermal conductivity. These
transport coefficients are given in terms of the usual Green-Kubo
formulae where the equilibrium averages are evaluated at the
instantaneous values of the temperature and chemical potential fields.

The drift term $K_{\bf r}^m[\tilde{\alpha}]$ in (\ref{deffunc}) is
obtained by collecting the results for the reversible part $v^m_{\bf
r}[\tilde{\alpha}]$ in (\ref{leres}), for the kinetic coefficient
(\ref{q2b}) and for the thermodynamic forces (\ref{intens0}). We will
have

\begin{eqnarray}
K_{\bf r}^0[\tilde{\alpha}]
&=& -\partial_{\bf r} \!\cdot\! {\bf g}_{\bf r}
\nonumber\\
{\bf K}_{\bf r}[\tilde{\alpha}]&=&-\partial_{\bf r} \!\cdot\! 
\left[{\bf v}_{\bf r}{\bf v}_{\bf r}\rho_{\bf r}+p_{\bf r}{\bf 1}-
2\eta\dot{\gamma}
-(\zeta-\frac{2}{3}\eta){\rm tr}[\dot{\gamma}]{\bf 1}\right]
\nonumber\\
K_{\bf r}^4[\tilde{\alpha}]&=&-\partial_{\bf r} \!\cdot\! \left[
{\bf v}_{\bf r}(e_{\bf r}+p_{\bf r})
-2\eta{\bf v}_{\bf r}\!\cdot\!\dot{\gamma}
-(\zeta-\frac{2}{3}\eta){\rm tr}[\dot{\gamma}]{\bf v}_{\bf r}\right.
\nonumber\\
 &-&\left.\kappa\nabla T_{\bf r}\right]
\label{drift0}
\end{eqnarray}
where
$\dot{\gamma}\equiv\frac{1}{2}\left[\nabla{\bf v}
+(\nabla{\bf v})^T\right]$
is the strain rate tensor. 

It is shown in Ref. \cite{zub83} that the term of $K_{\bf
r}^m[\tilde{\alpha}]$ in (\ref{deffunc}) vanish for the case of
hydrodynamic variables. This is
\begin{equation}
\int d{\bf r}'\frac{\delta}{\delta \tilde{\alpha}^{m'}_{{\bf r}'}}
\zeta^{mm'}_{{\bf r}{\bf r}'}[\tilde{\alpha}]=0
\label{i=s}
\end{equation}
This is a consequence of the locality in space assumption and has the
very important consequence that the It\^o and Stratonovich forms of the
FPE  coincide \cite{saa82}. In this way, the
resulting SDE can be interpreted in both senses with identical results.

The diffusion matrix $D^{mm'}_{{\bf r}{\bf r}'}$ (\ref{deffunc}) is
given by

\begin{eqnarray}
D^{\alpha\beta}_{{\bf r}{\bf r}'}
&=&-\partial_{{\bf r}_\mu}\partial_{{\bf r}_\nu }
\delta({\bf r}-{\bf r'})G^{\alpha\mu\beta\nu}_{\bf r}
\nonumber\\
D^{\alpha 4}_{{\bf r}{\bf r}'}
&=&-\partial_{{\bf r}_\mu}\partial_{{\bf r}_\nu }
\delta({\bf r}
-{\bf r'})G^{\alpha\mu\beta\nu}_{\bf r}{\bf v}^\beta_{\bf r}
\nonumber\\
D^{4 4}_{{\bf r}{\bf r}'}
&=&
-\partial_{{\bf r}_\mu}\partial_{{\bf r}_\nu}\delta({\bf r}-{\bf r'})
\left[2T^2({\bf r})\kappa({\bf r})\delta_{\mu\nu}\right.
\nonumber\\
&+&\left.
G^{\alpha\mu\beta\nu}_{\bf r}{\bf v}^\alpha_{\bf r}
{\bf v}^\beta_{\bf r}\right] 
\label{diff3}
\end{eqnarray}
where we have introduced

\begin{eqnarray}
G^{\alpha\mu\beta\nu}_{\bf r}&\equiv&
L^{\alpha\mu\beta\nu}_{\bf r}
+L^{\beta\mu\alpha\nu}_{\bf r}
\nonumber\\
&=&
2T({\bf r})\eta({\bf r})\delta_{\alpha\beta}\delta_{\mu\nu}
\nonumber\\
&+&T({\bf r})\left(\frac{1}{3}\eta({\bf r})+\zeta({\bf r})\right)
\left[\delta_{\alpha\mu}\delta_{\beta\nu}
+\delta_{\alpha\nu}\delta_{\beta\mu}\right]
\label{defg}
\end{eqnarray}
which is a tensor symmetric in $\alpha,\beta$ and $\mu,\nu$.

Now that we have the drift term $K_{\bf r}^m[\tilde{\alpha}]$ and the
diffusion matrix $D_{{\bf r}{\bf r}'}^{mm'}[\tilde{\alpha}]$ we could
write down the FPE (\ref{fpdens}) explicitly.

\section{SDE for simple fluids}

The main difficulty in writing down the SDE associated to the FPE for
hydrodynamic fields is that we have to compute the square root in
matrix sense of the diffusion matrix $D$ in (\ref{q2b}). This is not
trivial because this is a $5\times5$ matrix (rather, a $4\times4$
because it has the first column and first row equal to zero) that as
elements has continuous matrices with indices ${\bf r}{\bf
r}'$. However, one can resort to the physics of the problem to get a
hint on the solution. The hint comes from Eqn. (\ref{e9b}) that states
that the time derivative of conserved variables is in the form of the
divergence of a flux. This suggests that if we look at the noise term
(\ref{noiscont}) it should have the form of a divergence. A little of
reflection shows that we can look for a noise term of the form

\begin{eqnarray}
f^{\alpha}_{\bf r}(t)
&=&\partial_{{\bf r}^\beta}
\int d{\bf r}'\tilde{B}^{\alpha\beta\mu\nu}_{{\bf r}
{\bf r}'}[\tilde{\alpha}(t)]
\eta^{\mu\nu}_{{\bf r}'}(t)
\nonumber\\
f^{4}_{\bf r}(t)
&=&
\partial_{{\bf r}^\alpha}\int d{\bf r}'
\tilde{B}^{\alpha\beta}_{{\bf r}{\bf r}'}
\eta^{\beta}_{{\bf r}'}(t)
+
\partial_{{\bf r}^\alpha}\int d{\bf r}'{\bf v}_{{\bf r}'}^\beta 
\tilde{B}_{{\bf r}{\bf r}'}^{\alpha\beta\mu\nu}
\eta^{\mu\nu}_{{\bf r}'}(t)
\nonumber\\
\label{hint}
\end{eqnarray}
In these expressions, the white noise terms are given in terms of
independent increments of the Wienner process according to
(\ref{white}), this is

\begin{eqnarray}
\eta^{\mu\nu}_{\bf r}(t)
&\equiv& \frac {dW^{\mu\nu}_{\bf r}(t)}{\sqrt{v_{\bf r}}dt}
\nonumber\\
\eta^\mu_{\bf r}(t)
&\equiv& \frac {dW^{\mu}_{\bf r}(t)}{\sqrt{v_{\bf r}}dt}
\label{whi2}
\end{eqnarray}
The independent increments of the Wienner
process satisfy

\begin{eqnarray}
dW^{\mu\nu}_{\bf r}(t)&=&dW^{\nu\mu}_{\bf r}(t)
\nonumber\\
dW^{\mu\nu}_{\bf r}(t)dW^{\mu'\nu'}_{{\bf r}'}(t')&=&
[\delta^{\mu\mu'}\delta^{\nu\nu'}+\delta^{\nu\mu'}\delta^{\mu\nu'}]
\delta_{{\bf r}{\bf r}'}dt\quad \mbox{if $t=t'$} 
\nonumber\\
dW^{\mu\nu}_{\bf r}(t)dW^{\mu'}_{{\bf r}'}(t')&=&0
\nonumber\\
dW^{\mu}_{\bf r}(t)dW^{\mu'}_{{\bf r}'}(t')&=&
\delta_{\mu\mu'}\delta_{{\bf r}{\bf r}'}dt\quad\quad\quad 
\mbox{if $t=t'$} 
\label{wie1}
\end{eqnarray}

Whenever correlations of the white noise terms with other variables
have to be computed, one can use the fact that in It\^o sense the noise
and the variables are uncorrelated at the same time allowing for a
decoupling of the averages. Then one can use of the following
correlations for the white noise
\begin{eqnarray}
\langle \eta^{\mu\nu}_{\bf r}(t)\eta^{\mu'\nu'}_{{\bf r}'}(t')\rangle 
&=&
[\delta^{\mu\mu'}\delta^{\nu\nu'}+\delta^{\nu\mu'}\delta^{\mu\nu'}]
\delta({\bf r}-{\bf r}')\delta(t-t')
\nonumber\\
\langle \eta^{\mu\nu}_{\bf r}(t)\eta^{\mu'}_{{\bf r}'}(t')\rangle &=&0
\nonumber\\
\langle \eta^{\mu}_{\bf r}(t)\eta^{\mu'}_{{\bf r}'}(t')\rangle &=&
\delta_{\mu\mu'}\delta({\bf r}-{\bf r}')\delta(t-t')
\label{wie3}
\end{eqnarray}

Now, we make the ansatz that the matrices $\tilde{B}$ in (\ref{hint})
are local in space, this is

\begin{eqnarray}
\tilde{B}^{\alpha\beta\mu\nu}_{{\bf r}{\bf r'}}
&=&\tilde{B}^{\alpha\beta\mu\nu}_{{\bf r}}\delta({\bf r}-{\bf r'})
\nonumber\\
\tilde{B}^{\alpha\beta}_{{\bf r}{\bf r'}}
&=&\tilde{B}^{\alpha\beta}_{{\bf r}}\delta({\bf r}-{\bf r'})
\label{diag}
\end{eqnarray}
In addition, the form of the matrix $\tilde{B}$ must be simple
because, roughly speaking, its square must be like the matrix $G$ in
(\ref{defg}) which contains only products of Kronecker
deltas. Therefore we make the further assumption that the matrices
$\tilde{B}^{\alpha\beta\mu\nu}_{{\bf r}},\tilde{B}^{\alpha\beta}_{{\bf
r}}$ have the form

\begin{eqnarray}
\tilde{B}^{\alpha\beta\mu\nu}_{{\bf r}}
&=&a({\bf r})\delta_{\alpha\beta}\delta_{\mu\nu}
+b({\bf r})\left[\delta_{\alpha\mu}\delta_{\beta\nu}
+\delta_{\alpha\nu}\delta_{\mu\beta}\right]
\nonumber\\
\tilde{B}^{\alpha\beta}_{{\bf r}}&=&c({\bf r})\delta_{\alpha\beta}
\label{ans1}
\end{eqnarray}
where $a({\bf r}),b({\bf r}),c({\bf r})$ are functions to be specified
later. These expressions have the correct symmetry properties
for the matrices $\tilde{B}$'s. In the following, we will prove that
this ansatz is correct.

If we introduce the random stress tensor
$\tilde{\sigma}^{\alpha\beta}_{\bf r} (t)$ and random energy flux
$\tilde{\tau}^\alpha_{\bf r}(t)$ in such a way that
\begin{eqnarray}
f^\alpha_{\bf r}(t)&=&\partial_{{\bf r}_\beta}
\tilde{\sigma}^{\alpha\beta}_{\bf r} (t)
\nonumber\\
f^4_{\bf r}(t)
&=&\partial_{{\bf r}_\alpha}\tilde{\tau}^\alpha_{\bf r}(t)
\label{r2}
\end{eqnarray}
then we see from (\ref{hint}), (\ref{diag}), and (\ref{ans1}) that they
must have the form

\begin{eqnarray}
\tilde{\sigma}^{\alpha\beta}_{\bf r} (t)
&=&a({\bf r})\delta_{\alpha\beta}{\rm tr}[\eta_{\bf r}]
+ 2b({\bf r})\eta^{\alpha\beta}_{\bf r}
\nonumber\\
\tilde{\tau}^\alpha_{\bf r}(t)
&=&c({\bf r})\eta^{\alpha}_{\bf r}(t)+{\bf v}_{\bf r}^\beta(t)
\tilde{\sigma}^{\alpha\beta}_{\bf r} (t)
\label{form}
\end{eqnarray}

In order to determine the functions $a({\bf r}),b({\bf r}),c({\bf r})$
(and check that our ansatz is correct) we compute the correlation of
the noises using the expressions (\ref{r2}) with the definitions
(\ref{form}) and the properties (\ref{wie3}). We obtain

\begin{eqnarray}
\langle f^\alpha_{\bf r}(t)f^{\beta}_{{\bf r}'}(t')\rangle
&=&-\delta(t-t')
\nonumber
\\
&\times&\left[
\delta_{\alpha\beta}\nabla^2_{\bf r}\delta({\bf r}-{\bf r}')
\langle 4b^2({\bf r})\rangle\right.
\nonumber\\
&+&
\partial_{{\bf r}_{\alpha}}\partial_{{\bf r}_{\beta}}
\delta({\bf r}-{\bf r}')\langle 4b^2({\bf r})
+8a({\bf r})b({\bf r})
\nonumber
\\&+&\left. 6a^2({\bf r})\rangle\right]
\nonumber\\
\langle f^\alpha_{\bf r}(t)f^{4}_{{\bf r}'}(t')\rangle
&=&-\delta(t-t')
\nonumber
\\
&\times&
\left[
\nabla^2_{\bf r}\delta({\bf r}-{\bf r}')
\langle {\bf v}_{\bf r}^\alpha 4b^2({\bf r})\rangle
\right.
\nonumber\\
&+&
\partial_{{\bf r}_{\alpha}}\partial_{{\bf r}_{\beta}}
\delta({\bf r}-{\bf r}')
\langle{\bf v}^\beta_{\bf r}( 4b^2({\bf r})+8a({\bf r})b({\bf r})
\nonumber
\\
&+&\left.6a^2({\bf r}))\rangle\right]
\nonumber\\
\langle f^4_{\bf r}(t)f^{\alpha}_{{\bf r}'}(t')\rangle
&=&
\langle f^\alpha_{\bf r}(t)f^{4}_{{\bf r}'}(t')\rangle
\nonumber\\
\langle f^4_{\bf r}(t)f^{4}_{{\bf r}'}(t')\rangle
&=&-\delta(t-t')
\left[
\nabla^2_{\bf r}\delta({\bf r}-{\bf r}')
\langle c^2({\bf r})\rangle\right.
\nonumber\\
&+&\nabla^2_{\bf r}\delta({\bf r}-{\bf r}')
\langle {\bf v}_{\bf r}^2 4b^2({\bf r})\rangle
\nonumber\\
&+&
\partial_{{\bf r}_{\alpha}}\partial_{{\bf r}_{\beta}}
\delta({\bf r}-{\bf r}')
\langle{\bf v}^\alpha_{\bf r}{\bf v}^\beta_{\bf r}( 4b^2({\bf r})
\nonumber
\\
&+&\left.8a({\bf r})b({\bf r})+6a^2({\bf r}))\rangle
\right]
\label{noi2}
\end{eqnarray}

These expressions should coincide with (\ref{know1}) which can be
rewritten as

\begin{eqnarray}
\langle f^\alpha_{\bf r}(t)f^{\beta}_{{\bf r}'}(t')\rangle
&=&\delta(t-t')\langle D^{\alpha\beta}_{{\bf r}{\bf r}'}\rangle
\nonumber\\
\langle f^\alpha_{\bf r}(t)f^{4}_{{\bf r}'}(t')\rangle
&=&\delta(t-t')\langle D^{\alpha 4}_{{\bf r}{\bf r}'}\rangle
\nonumber\\
\langle f^4_{\bf r}(t)f^{4}_{{\bf r}'}(t')\rangle
&=&\delta(t-t')\langle D^{44}_{{\bf r}{\bf r}'}\rangle
\label{know2}
\end{eqnarray}
where the diffusion matrix is given in (\ref{diff3}).  After
substitution of (\ref{diff3}) into (\ref{know2}) one obtains

\begin{eqnarray}
\langle f^\alpha_{\bf r}(t)f^{\beta}_{{\bf r}'}(t')\rangle
&=&-\delta(t-t')
\left[
\delta_{\alpha\beta}\nabla^2_{\bf r}\delta({\bf r}-{\bf r}')
\langle 2T({\bf r})\eta({\bf r})\rangle
\right.
\nonumber\\
&+&\left.
\partial_{{\bf r}_{\alpha}}\partial_{{\bf r}_{\beta}}
\delta({\bf r}-{\bf r}')\langle2T({\bf r})\left(\eta({\bf r})/3
+\zeta({\bf r})\right)\rangle\right]
\nonumber\\
\langle f^\alpha_{\bf r}(t)f^{4}_{{\bf r}'}(t')\rangle
&=&-\delta(t-t')
\left[
\nabla^2_{\bf r}\delta({\bf r}-{\bf r}')
\langle {\bf v}_{\bf r}^\alpha 2T({\bf r})\eta({\bf r})\rangle
\right.
\nonumber\\
&+&
\partial_{{\bf r}_{\alpha}}\partial_{{\bf r}_{\beta}}
\delta({\bf r}-{\bf r}')
\nonumber
\\
&\times&\left.\langle{\bf v}^\beta_{\bf r}(2T({\bf r})
\left(\eta({\bf r})/3+\zeta({\bf r})\right))\rangle\right]
\nonumber\\
\langle f^4_{\bf r}(t)f^{\alpha}_{{\bf r}'}(t')\rangle
&=&
\langle f^\alpha_{\bf r}(t)f^{4}_{{\bf r}'}(t')\rangle
\nonumber\\
\langle f^4_{\bf r}(t)f^{4}_{{\bf r}'}(t')\rangle
&=&-\delta(t-t')
\left[
\nabla^2_{\bf r}\delta({\bf r}-{\bf r}')
\langle2T({\bf r})^2\kappa({\bf r})\rangle\right.
\nonumber \\
&+&\left.\nabla^2_{\bf r}\delta({\bf r}-{\bf r}')
\langle {\bf v}_{\bf r}^2 2T({\bf r})\eta({\bf r})\rangle
\right.
\nonumber\\
&+&
\partial_{{\bf r}_{\alpha}}\partial_{{\bf r}_{\beta}}
\delta({\bf r}-{\bf r}')
\nonumber
\\
&\times&\left.
\langle{\bf v}^\alpha_{\bf r}{\bf v}^\beta_{\bf r}(2T({\bf r})
\left(\eta({\bf r})/3+\zeta({\bf r})\right))\rangle
\right]
\label{noi3}
\end{eqnarray}
Compatibility between (\ref{noi2}) and (\ref{noi3}) then requires that

\begin{eqnarray}
4b^2({\bf r})&=& 2T({\bf r})\eta({\bf r})
\nonumber\\
3a^2({\bf r})+4a({\bf r})b({\bf r})+2b^2({\bf r})
&=& T({\bf r})\left[\frac{1}{3}\eta({\bf r})+\zeta({\bf r})\right]
\nonumber\\
c^2({\bf r})&=&2T^2({\bf r})\kappa({\bf r})
\label{comp}
\end{eqnarray}
We observe then that our ansatz for the matrices $\tilde{B}$ is correct
because this compatibility condition can be satisfied. The solution of
this system of equations is
\begin{eqnarray}
a({\bf r})+\frac{2}{3}b({\bf r})
&=&\pm \sqrt{\frac{1}{3}T({\bf r})\zeta({\bf r})}
\nonumber\\
b({\bf r})&=&\pm\sqrt{\frac{1}{2}T({\bf r})\eta({\bf r})}
\nonumber\\
c({\bf r})&=&T({\bf r})\sqrt{2\kappa({\bf r})}
\label{solu}
\end{eqnarray}
Therefore, the random stress tensor in
(\ref{form}) will have the following form
\begin{eqnarray}
\tilde{\sigma}^{\alpha\beta}_{\bf r}(t)  
&\equiv& \sqrt{2T({\bf r})\eta({\bf r})}
\left[\eta^{\alpha\beta}_{\bf r}(t)
-\delta^{\alpha\beta}
\frac{1}{3}\sum_\mu \eta^{\mu\mu}_{\bf r}(t)\right]
\nonumber\\
&\pm& \sqrt{3T({\bf r})\zeta({\bf r})}
\delta^{\alpha\beta}\frac{1}{3}\sum_\mu \eta^{\mu\mu}_{\bf r}(t)
\label{randoms0}
\end{eqnarray}
We have decomposed this tensor as a traceless part plus a diagonal
part. Both components are statistically independent and therefore the
sign $\pm$ is indifferent. Actually, we can compute the correlation of
the random stress tensor and the result is

\begin{eqnarray}
\langle\tilde{\sigma}^{\alpha\beta}_{\bf r}(t)  
\tilde{\sigma}^{\mu\nu}_{{\bf r}'}(t')  \rangle 
&=&\delta({\bf r}-{\bf r}')\delta(t-t')
\nonumber
\\
&\times&\left[ \langle 2T({\bf r})\eta({\bf r})\rangle 
[\delta^{\alpha\mu}\delta^{\beta\nu}+\delta^{\alpha\nu}\delta^{\beta\mu}]
\right.
\nonumber\\
&+&\left.
\langle 2T({\bf r})(\zeta({\bf r})-\frac{2}{3}\eta({\bf r}))\rangle 
\delta_{\alpha\beta}\delta_{\mu\nu}
\right]
\label{corr1c}
\end{eqnarray}
In the same fashion, we can define a random heat flux according to
\begin{equation}
\tilde{{\bf J}}^\alpha_{\bf r} \equiv
\tilde{\tau}^\alpha_{\bf r}(t)
-\tilde{\sigma}^{\alpha\beta}_{\bf r} (t){\bf v}_{\bf r}^\beta(t)
=
T({\bf r})\sqrt{2\kappa({\bf r})}\eta^\alpha_{\bf r}(t)
\label{heatf}
\end{equation}
This random heat flux is uncorrelated with the random stress tensor
and its autocorrelation is given by

\begin{equation}
\langle \tilde{{\bf J}}^\alpha_{\bf r}(t) 
\tilde{{\bf J}}^\beta_{{\bf r}'}(t') \rangle
=\langle 2T^2({\bf r})\kappa({\bf r})\rangle 
\delta_{\alpha\beta}\delta({\bf r}-{\bf r}')\delta(t-t')
\label{corr2b}
\end{equation}
The expressions (\ref{randoms0}),(\ref{heatf}) of the random stress
tensor and random heat flux in terms of the independent increments of
the Wienner process are the main results of this paper.

Now, we have all the elements to write the full Langevin equations
(\ref{cont8}).  After some rearrangement, the Langevin equations can
be cast in the form
\begin{eqnarray}
\partial_t\rho({\bf r},t)
&=&-\nabla\!\cdot\!{\bf v}({\bf r},t)\rho({\bf r},t)
\nonumber\\
\partial_t{\bf g}({\bf r},t)
&=&-\nabla\!\cdot\!\left[{\bf v}({\bf r},t){\bf g}({\bf r},t)
+\sigma({\bf r},t)\right]
\nonumber\\
\partial_te({\bf r},t)
&=&-\nabla\!\cdot\! \left({\bf v}({\bf r},t)e({\bf r},t)
+{\bf v}({\bf r},t)\!\cdot\!\sigma({\bf r},t)
+{\bf J}({\bf r},t)\right)
\nonumber\\
\label{bal}
\end{eqnarray}
The terms ${\bf v}\rho,{\bf v}{\bf g}, {\bf v}e$ represent the purely
convective part of the fluxes. The total stress tensor $\sigma$ and
total heat flux ${\bf J}$ are given by a systematic and a random part
\begin{eqnarray}
\sigma({\bf r},t) &=& \overline{\sigma}({\bf r},t)
+\tilde{\sigma}({\bf r},t)
\nonumber\\
{\bf J}({\bf r},t) &=& \overline{{\bf J}}({\bf r},t)
+\tilde{{\bf J}}({\bf r},t)
\end{eqnarray}
The systematic parts have the form of Newton's and Fourier's 
constitutive equations

\begin{eqnarray}
\overline{\sigma}&=& p{\bf 1}
-2\eta\left(\dot{\gamma}
-\frac{1}{3}{\rm tr}[\dot{\gamma}]{\bf 1}\right)
-\zeta{\rm tr}[\dot{\gamma}]{\bf 1}
\nonumber\\
\overline{{\bf J}}&=&-\kappa\nabla T
\label{constitutive}
\end{eqnarray}
where
$\dot{\gamma}\equiv\frac{1}{2}\left[\nabla{\bf v}
+(\nabla{\bf v})^T\right]$
is the strain rate tensor. The random parts are given by
(\ref{randoms0}),(\ref{heatf}).

\section{Discussion}

The Langevin equations for non-linear hydrodynamic are the same in
structure as the usual macroscopic non-linear hydrodynamic equations
except that they govern the instantaneous fluctuating values of the
hydrodynamic fields and they have, in addition, a random contribution
to the stress tensor and heat flux.  The form in which the thermal
noise is described here is similar to the Landau and Lifshitz theory
\cite{lan59}. The differences arise on the stochastic properties of
the random parts in these equations. The correlations of the random
stress tensor and heat flux, Eqn. (\ref{corr1c}),(\ref{corr2b}) are to
be compared with those given by Landau and Lifshitz \cite{lan59}:

\begin{eqnarray}
\langle\tilde{\sigma}^{\alpha\beta}_{\bf r}(t)  
\tilde{\sigma}^{\mu\nu}_{{\bf r}'}(t')  \rangle 
&=&\delta({\bf r}-{\bf r}')\delta(t-t')
\nonumber
\\
&\times&
\left[ 2T_{eq}\eta_{eq} \left(
\delta^{\alpha\mu} \delta^{\beta\nu}+
\delta^{\alpha\nu} \delta^{\beta\mu}\right)\right.
\nonumber
\\
&+&\left.
2T_{eq}(\zeta_{eq}-\frac{2}{3}\eta_{eq})
\delta_{\alpha\beta}\delta_{\mu\nu}
\right]
\nonumber\\
\langle \tilde{{\bf J}}^\alpha_{\bf r}(t)
 \tilde{{\bf J}}^\beta_{{\bf r}'}(t') \rangle
&=&2T^2_{eq}\kappa_{eq}
\delta_{\alpha\beta}\delta({\bf r}-{\bf r}')\delta(t-t')
\label{LL}
\end{eqnarray}
Here $T_{eq},\eta_{eq},\zeta_{eq},\kappa_{eq}$ are the equilibrium
values of the temperature and the transport coefficients. It is
apparent that the results (\ref{corr1c}),(\ref{corr2b}) will coincide
with (\ref{LL}) only if the system is at equilibrium and the
fluctuations can be considered small, in such a way that the decoupling
approximation
\begin{equation}
\langle T({\bf r})\eta({\bf r})\rangle =
\langle T({\bf r})\rangle\langle\eta({\bf r})\rangle = T_{eq}\eta_{eq}
\label{decoup}
\end{equation}
can be taken. 

Another point of discrepancy is that the random fluxes {\em are not
Gaussian}.  What is Gaussian is the Wienner process, but not the
product of the Wienner process with the temperature and the transport
coefficients. This result is consistent with that of Ref. \cite{saa82}
where the assumption of Gaussian random fluxes lead to unphysical
results. In particular, it was found that if the Einstein distribution
is assumed to be the equilibrium solution, then the viscosities were
inversely proportional to the temperature and that the heat
conductivity was inversely proportional to the temperature
squared. This dependences are not observed in real
systems. Conversely, one could obtain physically acceptable transport
coefficients by relaxing the requirement that the Einstein
distribution is the equilibrium, which is equally unacceptable.  A
full discussion of this point has been made in Ref. \cite{zub83}.

The ``size of the fluctuations'' is what determines the validity of
the decoupling approximation (\ref{decoup}) and we comment on this
point now. We note that despite of the continuum appearance of the
Langevin equations (\ref{bal}) there is an implicit dependence of the
random parts on the coarse graining, i.e. on the volume of the
cells. This is apparent in the definition (\ref{whi2}) of the noise
appearing in (\ref{randoms0}),(\ref{heatf}).  In principle, different
coarse-graining procedures (different discretization with distinct
typical volumes $v_{\bf r}$ of the cells) will leave the systematic
part of the equations invariant, but will change the intensity of the
noise terms. The $1/\sqrt{v_{\bf r}}$ dependence is consistent with
the amplitude of the fluctuations encountered in equilibrium ensemble
theory. Therefore, the question of whether the fluctuations are large
or small is a matter of the level of description used. Fine
resolutions of the hydrodynamic fields in which the coarse graining
cells are small will produce large relative fluctuations in the
hydrodynamic variables. The existence of characteristic length scales
imposed on the fluid is what dictates the level of coarsening
required. In the case of colloidal particles, it is observed
experimentally that submicronic particles diffuse whereas the
diffusion for larger particles is small and negligible. Therefore, in
order to ``resolve'' submicronic particles one needs a fine level of
coarse graining such that the fluctuations are relevant.

The Landau and Lifshitz fluctuating hydrodynamics arises as the
small noise expansion \cite{gar83} of the SDE obtained in this
paper. This can be seen by writing schematically the SDE
(\ref{cont8}) with (\ref{diag}) as 

\begin{equation}
d\tilde{\alpha} = K[\tilde{\alpha}] dt 
+\epsilon \tilde{B}[\tilde{\alpha}]dW(t)
\label{esc}
\end{equation}
where $\epsilon=1/\sqrt{v_{\bf r}}$. In a perturbative expansion in
$\epsilon$ one assumes that the solution of (\ref{esc}) can be written
as $\tilde{\alpha}(t)=\tilde{\alpha}_0(t)+\epsilon\tilde{\alpha}_1(t)
+\epsilon^2\tilde{\alpha}_2(t)+\cdots$. By substitution in (\ref{esc})
and cancelling order by order one obtains

\begin{eqnarray}
d\tilde{\alpha}_0(t) &=& K[\tilde{\alpha}_0(t)] dt 
\nonumber \\
d\tilde{\alpha}_1(t) &=& 
K'[\tilde{\alpha}_0(t)]\tilde{\alpha}_1(t) dt 
+ \tilde{B}[\tilde{\alpha}_0(t)]dW
\nonumber \\
d\tilde{\alpha}_2(t) &=& 
\left[K'[\tilde{\alpha}_0(t)]\tilde{\alpha}_2(t) dt 
+\frac{1}{2}K''[\tilde{\alpha}_0(t)]\tilde{\alpha}_1(t)\right] dt 
\nonumber\\
&+& \tilde{B}'[\tilde{\alpha}_0(t)]\tilde{\alpha}_1(t)dW(t)
\label{esc2}
\end{eqnarray}
where the primes denote derivatives. The first equation for the
dominant contribution $a_0(t)$ to the solution of the SDE constitutes
the deterministic non-linear hydrodynamic equations with fluctuations
neglected. The next equation for the contribution $\alpha_1(t)$ is
equivalent to linearize the equations of hydrodynamics around the
deterministic solution and add a Gaussian white noise of amplitude $
\tilde{B}[\tilde{\alpha}_0]$. At long times when the system has
relaxed to equilibrium, these equations are the Landau and Lifshitz
linear equations governing the equilibrium hydrodynamic fluctuations
\cite{lan59}. Higher order contributions $\tilde{\alpha}_2,\ldots$
are usually not considered, although, in principle, they can be
computed systematically \cite{gar83}.

\section*{Acknowledgments}
I am very much indebted to M.A. Rubio, I. Z\'{u}\~niga, F.J. de la
Rubia and H.C. \"Ottinger for their comments and suggestions during
the elaboration of this work. This work has been partially supported
by a DGICYT Project No PB94-0382.

\end{document}